# Hot carrier distribution engineering by alloying: picking elements for the desired purposes

Matej Bubaš and Jordi Sancho-Parramon*

*Ruđer Bošković Institute, Bijenička cesta 54, Zagreb 10000, Croatia*

E-mail: jsancho@irb.hr

**Abstract**

Metal alloys hold the promise of providing hot carrier generation distributions superior to pure metals in applications such as sensing, catalysis and solar energy harvesting. Guidelines for finding the optimal alloy configuration for a target application require understanding the connection between alloy composition and hot carrier distribution. Here we present a DFT-based computational approach to investigate the photo-generated hot carrier distribution of metal alloys based on the joint density of states and the electronic structure. We classified the metals by their electronic structure into closed d-shell, open d-shell, p-block and s-block elements. It is shown that combining closed d-shell elements enables modulating the distribution of highly energetic holes typical of pure metals but also leads to hot carrier production by IR light excitation and the appearance of highly energetic electrons due to band folding and splitting. This feature arises as an emergent property of alloying and is only unveiled when the hot carrier distribution computation takes momentum conservation into account. The combination of closed d-shell with open d-shell elements allows an abundant production of hot carriers in a broad energy range, while alloying a closed d-shell elements with an s-block element opens the door to hot electron distribution skewed toward high energy electrons. The combination of d-shell with p-block elements results in moderate hot carrier distribution whose asymmetry can be tuned by composition. Overall, the obtained insights that connect alloy composition, band structure and resulting carrier distribution provide a toolkit to match elements in an alloy for the deliberate engineering of hot carrier distribution.

# Introduction

Hot carriers (HCs) generated by plasmon decay in metal nanoparticles have recently been a subject of intense research due to the promise they hold in several research areas. In sensing utilizations, substantial advancements have been made for improving the performance of electrochemical sensors,[1] as well as photodetection,[2] enabling the design of novel optoelec-tronic devices.[3] In solar energy harvesting, HC generation can contribute through increasing efficiency for standard semiconductor-based photovoltaics alone,[4,5] in combination with organic photovoltaics,[6,7] and perovskite solar cells.[8] Current contribution of HCs to the field of catalysis is also rather compelling, as their unique properties facilitate previously unattainably low temperature catalysis,[9] enhance reaction selectivity, and unlock completely novel reaction pathways.[10,11] This enables catalysis of reactions that are both thermodynamically and/or kinetically quite unfavorable and thus complicated to achieve, such as Haber-Bosch process, one of the most important industrial procedures in modern times.[12]

Progress in the field of plasmon-based HC generation has led to different strategies and ways of HCs utilization, making use of HC generation in nanostructures composed of pure plasmonic metals,[13] in combination with semiconductors,[14,15] and also with good catalytic (but poor plasmonic) metals in antenna-reactor systems.[10] Despite the fast rate of exploration and discovery of new HC-based applications there is a need for understanding how to control and tune the HC generation to suit the wanted purpose, such as injection in a semiconductor or catalysis of a particular reaction. A crucial property of interest then becomes the energy distribution of HCs created in a plasmonic nanostructure. These highly energetic electrons and holes are short lived, complicating experimental determination of their energy distribution,[16] thus making density functional theory (DFT) calculations a popular way to tackle the problem. Several approaches, such as changing the size and shape of the nanoparticle, have been studied as a way of tuning the energy distribution of the HCs for the desired purposes.[17] Using different materials (usually gold or silver) to obtain distinct distributions has also been explored,[13,18–20] but due to the small number of good plasmonic metals hot carrier distribution is largely limited by their electronic structure.

In our previous research, we used DFT to explore how the electronic structure, as well as various optical and plasmonic properties, can be deliberately tuned by alloying. Furthermore, we identified new interband transitions that are an emergent property of alloying.[21] Based on the obtained knowledge, it follows that alloying could also affect the HC generation, and that DFT could provide insight into how to modulate or even tune it.

In this work, we present a straightforward DFT-based method suitable for high throughput calculations needed to study many alloyed systems, but strict enough to properly identify effects of alloying on hot electron and hot hole energy distribution. We explore how alloying different elements, and in different ratios, affects the hot carrier energy distribution. Finally, based on the findings, we identify strategies that enable engineering of hot carrier generation for desired purposes.

# Methods

GPAW package[22,23] and Atomic simulation environment[24] were used for DFT calculations. In all cases plane waves were used as a basis set and Brillouin zone was sampled based on Monkhorst-Pak grid.

For structure optimization Fermi-Dirac smearing was used, with a width of 0.1 eV. Unit cells of FCC-like alloys without experimentally available crystal structures were optimized using PBEsol functional[25] with k-point density of 15 k-points per reciprocal Å. When possible, experimentally determined crystal structure was used. Plane wave energy cutoff was set to 500 eV to ensure that Pulay stress was negligible for the results of the optimization, and the geometry optimization cutoff of 0.3 eV per Å was used for forces acting on all individual atoms.

Density of states calculations were performed using a 400 eV cutoff and a k-point density of 30 k-points per reciprocal Å.

Hot carrier energy distributions obtained from the density of states calculations were obtained by the method outlined in the works by White and Catchpole[26] and updated by Gong and Munday to include Fermi-Dirac distribution.[27] Essentially, the hot electron energy distribution is computed by neglecting momentum conservation in the photoexcitation transition probability and multiplying the density of filled states with the density of unfilled states that are separated by the photon energy and with the corresponding occupancy factors of initial and final states. Hot hole distribution is obtained in a similar way taking into account the energy of generated holes -instead of generated electrons- with respect to the Fermi level.[28]

In the approach here presented, hot carrier energy distributions are obtained taking into account momentum conservation and thus reflect the effect of direct transitions only. The method is based on an explicit calculation of the joint density of states[29] and the hot carrier distributions are computed by adding up all possible transitions between initial and final states with the same wave vector and an energetic separation equal to the photon energy. Therefore it is necessary to consider a k-dependent electronic structure. In order to prevent unnecessary band folding and simultaneously maximize the computational efficiency, before the calculation each unit cell is reduced to a primitive unit cell (minimal amount of atoms) representing the same lattice using the *pyspglib* module. Afterwards, for each k-point in the Brillouin zone eigenstates are calculated, with their respective occupancies. Since Fermi-Dirac distribution in this case directly influences the results of the hot carrier energy distribution, the width is not set to 0.1 eV as it is usually done to ensure fast convergence, but to 25.7 meV. which corresponds to approximately room temperature. Density of k-points is set to 50 reciprocal Å, to ensure sufficiently good resolution of the electronic structure. Several empty bands are included in the calculation and plane wave cutoff is set to 450 eV. The same parameters are used for regular band structure calculations. Excitations are counted between all states separated by photon energy broadened by ±8 meV. For each initial state a hot hole generation is counted and weighted by the respective occupancies of the initial and final states. In the same way, a hot electron generation is counted at the energy of the final state. Finally, a histogram of hot electron and hot hole distributions is created and represented using Gaussian convolution for smoothing.

# Results and discussion

## Importance of momentum conservation in calculations

Since the number of electrons needed to properly describe the metal nanoparticle is substantial, DFT calculations on such systems are very demanding and have insofar been performed only for nanoclusters smaller than 4 nm. Considering the large composition space to explore - several elements in a multitude of ratios - high-throughput methods, such as the one we present here, are more appropriate for alloyed systems. The efficiency is ensured by using bulk systems to obtain the information regarding the electronic structure. The results should be well translatable to thin films and medium to large nanoparticles, with limitations of such an approach having notable importance only for the nanoparticles smaller than 10 nm.[18,30]

We focus only on the direct transitions since they are more sensitive to the electronic structure changes and dominate phonon-assisted transitions above the interband threshold. In our previous research we have shown that alloying good plasmonic metals such as Au, Ag and Cu with one another unlocks new direct sp-sp transitions, which affects the optical losses in the low energy range. Figure 1 illustrates the importance of taking conservation of momentum into account when considering alloy systems. The results show the initial hot carrier energy distributions for Cu and $Cu_3Au$ alloy upon illumination with 2.8 eV light estimated with our method and the density of states (DOS)-based method that neglects the conservation of momentum.

$Cu_3Au$ alloy with an FCC-like $L1_2$ crystal structure was chosen as a model system since this ordered, intermetallic phase has been experimentally produced and characterized, so the computational picture closely reflects a realistic picture of an experimentally available system. Thus, the discrepancy between the idealized computational representation and the real system is minimized, and so is the influence of such discrepancy on the resulting hot carrier distributions.

It is visible that results obtained by a DOS-based method for a pure metal and an alloy can barely be distinguished. In contrast, the results obtained by our method show that the hot carrier distribution of an alloy system obtains an additional feature that is not present for pure Cu. In addition to a narrow distribution of highly energetic hot holes and low-energy hot-electrons which are present for the pure Cu, the $Cu_3Au$ alloy also produces a high-energy hot electron "tail", and the corresponding lower energy holes.

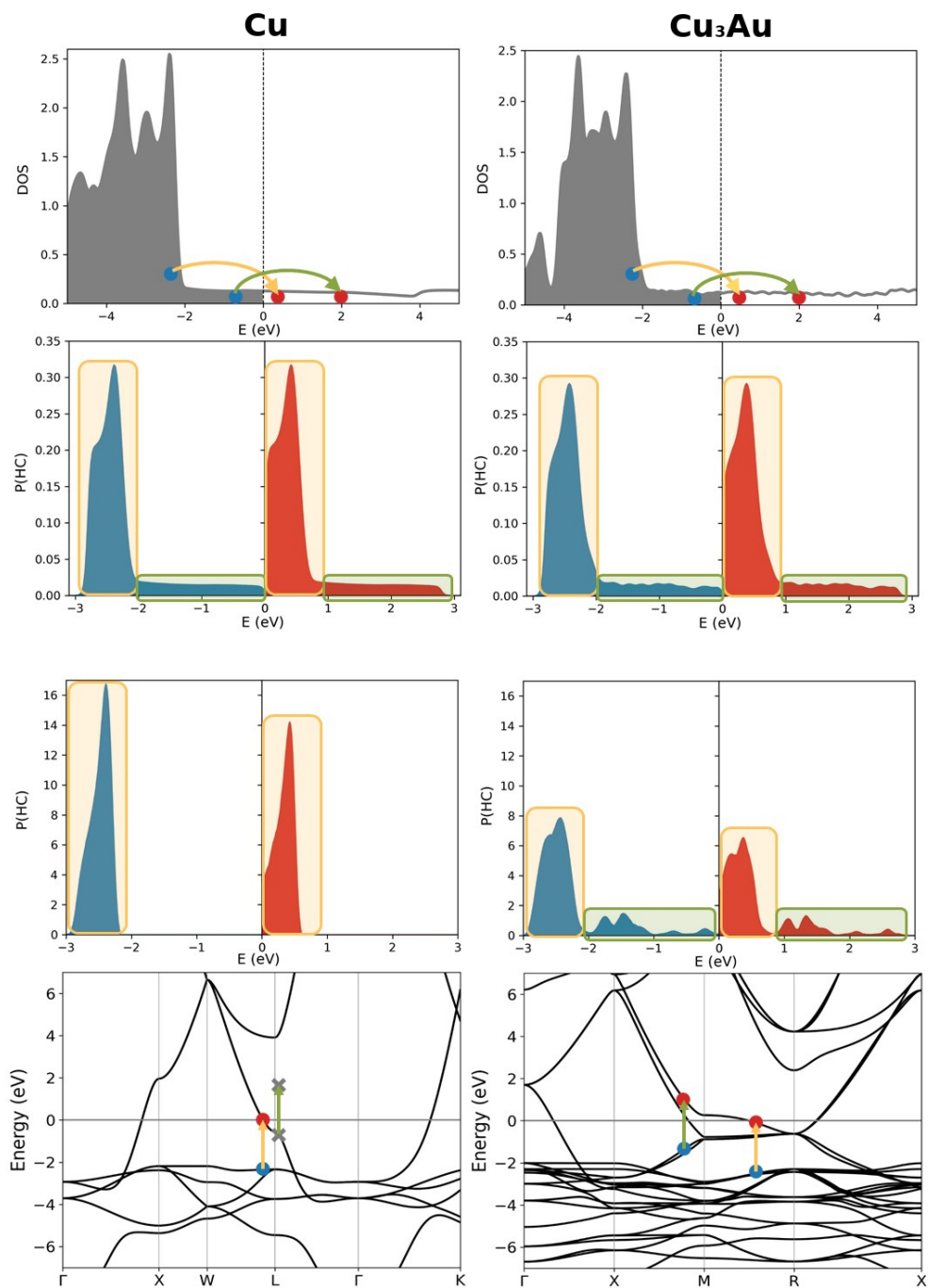

Figure 1: Comparison of hot carrier energy distributions obtained by a DOS-based method that neglects momentum conservation (top two rows) and our method that takes it into account (bottom two rows). Yellow arrows show the d-sp transitions, with the corresponding part of the hot carrier distribution shaded yellow. Green arrow shows sp-sp transitions with green shading showing the corresponding parts of the hot carrier distribution. Hot hole distributions are shown in blue and hot electron distributions in red.

## Effects of unit cell configuration and complexity (order-disorder influence)

Since most of the alloy systems of interest do not have a simple crystal structure or a high level of ordering, we attempted to grasp the order-disorder influence by changing the number of atoms and the configuration of the alloy unit cell for a given composition. For that purpose we compared the hot carrier distributions obtained using a highly symmetrical 2-atom AuCu unit cell with an 8-atom $Au_{0.5}Cu_{0.5}$ alloy representation constructed using a special quasirandom structure (SQS) Figure 2. SQS construction is designed to best approximate the disordered structure by using an ordered representation of a given size.[31] The difference in hot carrier generation is reflected in subtle feature discrepancies - most notably an overall smoother distribution for the SQS system. For the rest of the calculations performed in this work we take advantage of this notion, by representing the alloy systems by the simplest configuration of an appropriate composition, thus maximizing the efficiency of the calculations. The reasons for the observed behavior are further discussed in the Supplementary material, section "Effects of configuration on hot carrier distribution".

## Effects of combining closed-shell (filled d-orbitals) metals:

### a) obtaining a hot electron high energy tail due to emergent interband transitions

By unlocking new direct transitions upon alloying, hot electron distribution may be extended up to the photon energy. The reason for that can be rationalized by the changes in the band structure upon alloying. Alloyed structure leads to translational symmetry breaking of the unit cell which results in band folding. Upon folding, states of different energy in a single band end up at the same k-point (vertically above one another), which allows direct transitions between them. Besides that, band degeneracy is broken due to the hybridization with the orbitals of two different elements, which causes "band splitting" - loss of degeneracy resulting in several closely spaced and often near parallel bands.[21,32]

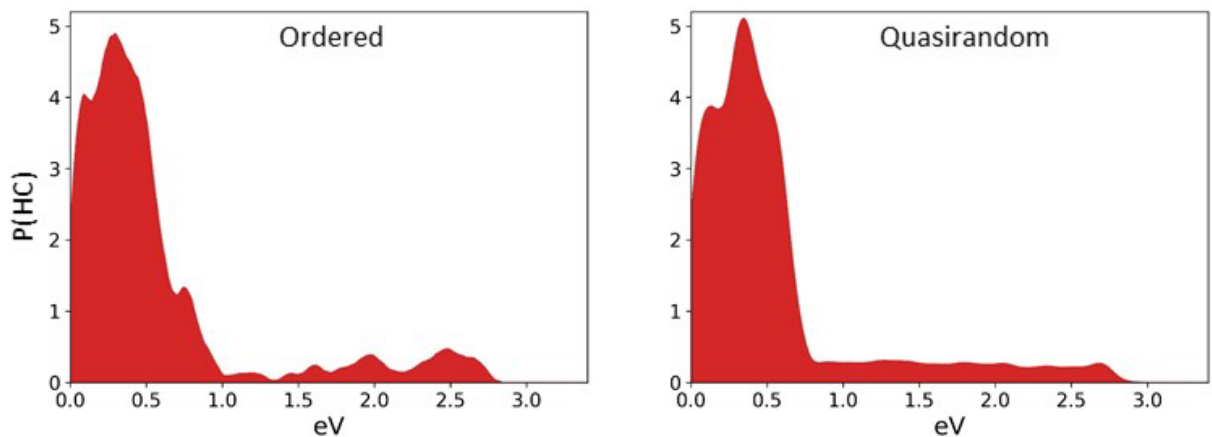

Figure 2: Hot electron generation probability for ordered AuCu alloy calculated using a simple 2-atom unit cell (left) and using a special quasirandom structure representing disordered alloy (right).

Since electronic structures of noble metals commonly used in plasmonics such as Ag, Au, and Cu are relatively similar (low lying d-bands, far below the Fermi level, sp-bands at similar positions) their alloying does not change the DOS a lot (an example shown in Figure 1). However, hot carrier generation is still modulated in different ways, through band splitting and folding, which leads to generation of hot electrons even at very low energies (IR range) and to a continuous distribution of electrons and holes with energies up to the photon energy above and below the Fermi level.

The presented results indicate that alloying leads to expanding the functionality of closed d-shell metals. Due to their electronic structure, namely high density of d-states below the Fermi level, photons in the visible range mostly excite direct transitions from low-lying d-bands to the vicinity of the Fermi level. This results in generation of relatively low-energy hot electrons. Generation of higher energy electrons in these metals is thus dependent on less tunable and less efficient phonon-assisted transitions, and on surface-assisted (also sometimes called geometry-assisted) transitions which quickly become less prominent with increasing nanoparticle size.[18]

However, alloying noble metals remedies this issue and enables the production of high-energy electrons. This especially affects larger nanoparticles because of the lower proportion of surface-assisted transitions, and provides them with better electron-injection properties. Production of high energy hot electrons increases the probability of injection over the Schottky barrier both by providing a larger proportion of electrons with sufficient energy, and due to the probability of injection of sufficiently energetic hot electrons increasing with the excess kinetic energy, by expanding the emission cone.[26] Additionally, chemical reactions that require electron injection into a high-energy orbital also benefit from a high-energy hot electron tail. It is worth noting that high-energy tail also positively affects uses which require only lower-energy electrons, due to hot carrier multiplication.[2,33] The number of hot electrons can be multiplied, as one high-energy electron can create several other hot electrons by transferring a part of its energy through electron scattering. Therefore, an initial high-energy tail can have an outsized contribution to the number of low-energy hot electrons after scattering.

**b) Modifying production of abundant deep hot holes and shallow hot electrons by d-band tuning**

Alongside introducing a "tail" in the hot carrier distribution, alloying of two closed d-shell elements leads to a shift in d-band position which can also be used to deliberately affect hot carrier distribution. If the photon energy $(E_{ph})$ is insufficient to excite electrons from the d-band to the Fermi energy $(E_F)$, only a continuous distribution of hot electrons (from $E_F$ to $E_F + E_{ph}$) and hot holes (from $E_F$-$E_{ph}$ to $E_F$) due to sp-sp transitions will be present. However, if the d-band is high enough, the photons of the same energy would be sufficient to excite electrons from the d-band to the sp-band above the Fermi level, resulting in abundant shallow (low energy) hot electrons and deep (high energy) hot holes (Figure 3). Alloying good plasmonic metals with a closed d-shell, such as Au and Ag, results in a relatively smooth shift of the d-band edge, between lower-lying Ag d-bands and higher-lying Au d-bands. This way, by choosing the appropriate composition, the hot carrier distribution feature coming from d-bands can be introduced (Shown on Figure 1) and tuned to the preferred degree as results visible on Figure 3 for alloys with 50% or larger Au content. For illustration purposes, the photon energy on Figure 3 is kept constant (2.8 eV, between plasmon resonances of Au and Ag spherical nanoparticles), regardless of the changing plasmon resonance with composition change that would occur in nanostructures. Since the hot electron and hot hole distributions obtained by this method are nearly perfectly symmetrical, only the hot electron distributions are shown in the main body of this work, with hot hole distributions presented in Supplementary Information on Figure S1.

In contrast to hot electron distribution in noble nanoparticles, most of the generated hot holes are highly energetic, which seems to indicate that the pure noble metal nanoparticles should be suitable for hole injection into semiconductors or molecules. Introducing abundant hot hole generation is especially important since hot hole mobility is adversely affected by their high effective mass, in part due to generally low valence band curvature.

**c) Hot carrier production by direct transitions extending to IR range**

In pure metals the interband transition threshold is determined by the photon energy required to excite an electron from the valence d-band to the conduction sp-band. As seen above, the d-sp transition threshold can be changed, primarily by the d-band tuning by alloying. However, due to the conduction band folding and splitting, new allowed sp-sp transitions require lower energy photons and effectively reduce the direct transition threshold quite substantially. Consequently, hot carrier generation as a result of direct transitions extends well into the IR range for such alloys. This is especially important since, in the absence of direct transitions, resistive losses which do not result in hot carrier generation make up to 50% of all losses.[18] However, introducing direct transitions as a competing loss mechanism should increase the efficiency of hot carrier generation in the energy range below the d-sp transition threshold. Such an observation has been made by Brown et al. for hot carrier generation in pure aluminum, with estimated 25% carriers generated by direct transitions even at very low energies. Because of direct transitions at all photon energies and the lowest proportion of resistive losses (when compared to Au, Ag and Cu) they have highlighted Al as a potentially great material for plasmonic applications.[18] The continuous production of HCs due to interband transitions is enabled by a band crossing near the W point in the band structure of Al.[30] In fact, a connection to the hot carrier production in alloys can be made since a very similar pattern in the band structure is produced by band folding Figure 1, thus extending the range of direct transition hot carrier production as seen on the Figure 3. It follows that alloys of closed d-shell metals exhibit hot carrier generation characteristics similar to that of Al and thus might be especially desirable materials for hot carrier generation in the low energy range.

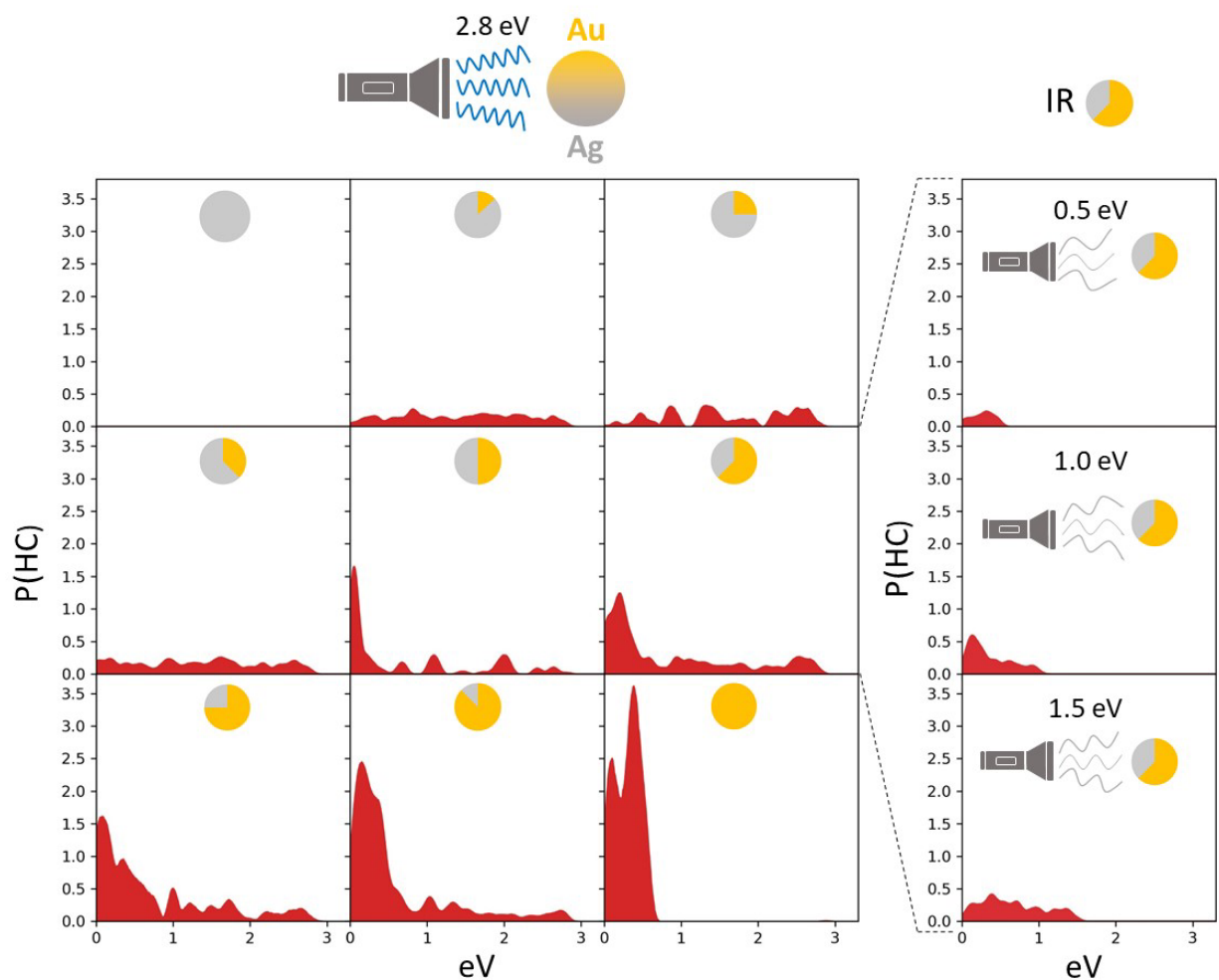

Figure 3: Hot electron energy distribution due to excitation with 2.8 eV light for Au-Ag alloy systems with the composition changing in 12.5% increments from pure Au to pure Ag (three columns on the left). The column on the right shows that hot electron distributions can be obtained by IR light excitations for a fixed alloy composition $Au_{0.625}Ag_{0.375}$, far below the nominal d-sp direct transition onset.

## Influence of combining closed and open d-shell elements in an alloy - enabling abundant and continuously distributed hot carrier generation due to high d-bands

A more significant influence can be achieved by alloying common plasmonic metals with Pd or Pt. These metals have an open d-shell, which adversely affects their performance as plasmonic metals due to plasmon quenching, but positively affects their catalytic properties. Due to their catalytic properties, recently these metals have been utilized in combination with good plasmonic metals in antenna-reactor configurations and in alloys.[34] Therefore, we have investigated how high-lying d-states in such metals affect the hot carrier generation upon alloying with low-lying d-state metals, since combining Pd with Au quickly introduces a high density of filled d-states, up to the Fermi level, and even crossing it.[35]

Due to the larger density of filled states that can be excited with the available photon energy, the high energy tail of hot electron distribution is much more prominent (Figure 4. Therefore, alloying good catalytic and good plasmonic noble metals provides a new degree of freedom for tuning the hot carrier distribution. Introducing high-lying d-states not only provides a new channel for plasmon decay by direct transitions but also significantly changes the DOS near the Fermi level. This could serve as a knob to adjust the extension of the high-energy tail of hot electron distribution to enhance electron-injection potential, as it would translate to a much larger hot electron generation probability at high energies given the same field enhancement in the nanoparticle.

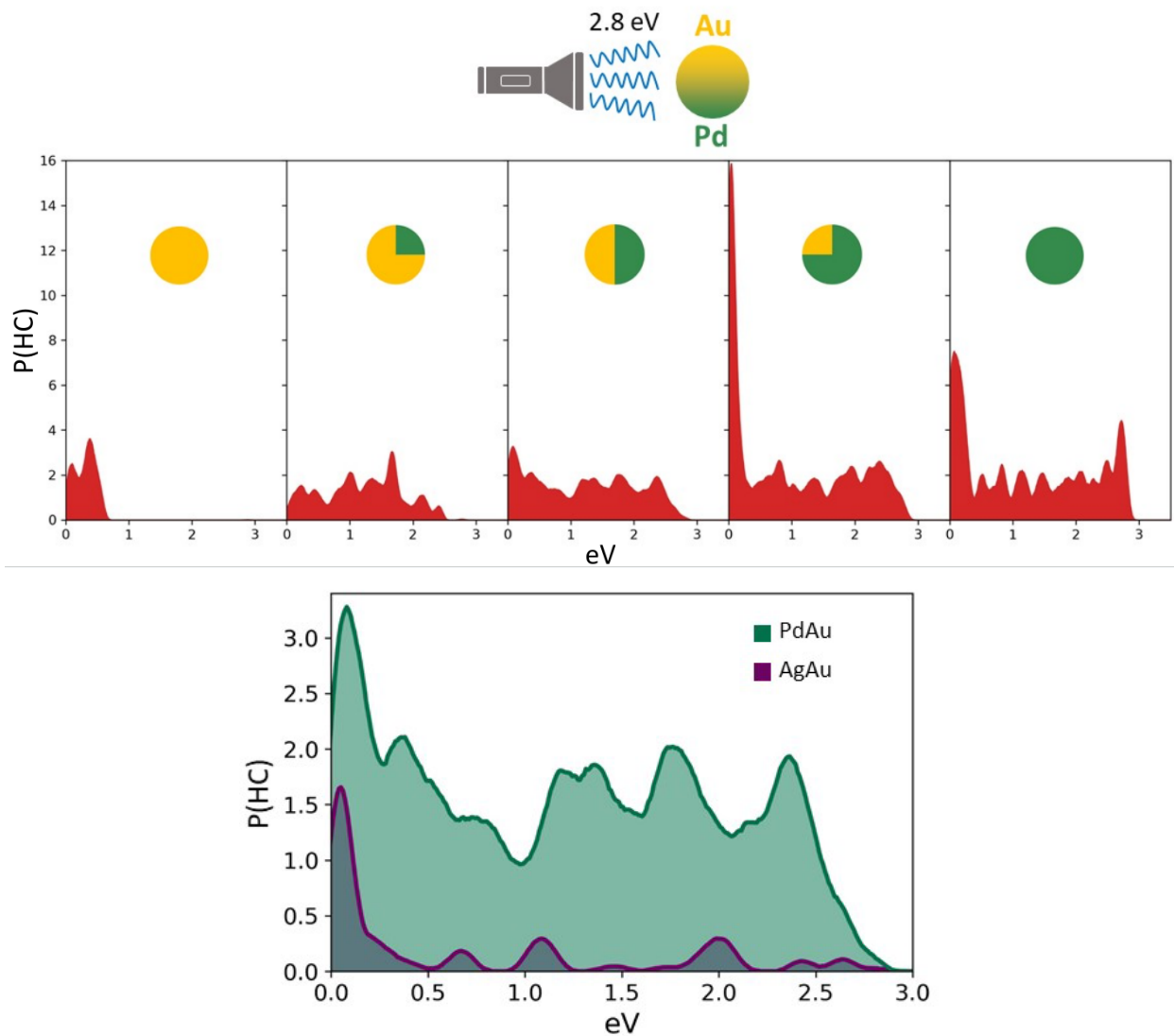

Figure 4: Top: Hot electron energy distributions obtained by a 2.8 eV excitation of Pd-Au systems ranging in composition from pure Pd to pure Au. Thte composition is varied by 25% increments. Bottom: Comparison of Hot electron energy distributions obtained after 2.8 eV excitation of PdAu system and AgAu system.

To illustrate the magnitude of (especially high energy) hot electrons introduced by alloying of Au with Pd, in Figure 4 we compare the hot electron energy distribution for $Au_{0.5}Pd_{0.5}$ alloy with that of the $Au_{0.5}Ag_{0.5}$ alloy. It is visible that above 0.5 eV Au-Pd alloy results in the distribution about an order of magnitude larger than for Au-Ag.

However, this is precisely the caveat of enabling many allowed direct transitions by alloying. Although alloying can be a way of introducing direct transitions close to the surface plasmon frequency, thus tuning the hot carrier generation distribution, such direct transitions also serve as an efficient plasmon decay pathway.[36] The resulting quenching of the plasmon leads to lower field enhancement, reducing the amount of light absorbed by the nanoparticle, and the consequent hot carrier generation. Therefore, there is a trade-off between the absolute amount of light absorbed and the efficiency of converting the energy of the absorbed light into hot carriers (manuscript on that topic is currently being prepared for publishing). Additionally, the proportion of "useful" hot carriers also has to be considered since, in the example of electron injection over the Schottky barrier, only the proportion of hot electrons with energies sufficient to traverse the barrier will be of importance.

## Influence of mixing s-block elements with filled d-shell elements - increasing the high-energy end of the hot electron energy distribution

An interesting possibility of hot carrier generation tuning by alloying comes from combining standard, filled d-shell plasmonic metals with s-block elements. Recently, s-block elements such as Na, K and Mg, have been pointed out as potentially promising plasmonic metals.[37,38] However, their experimental utilization is largely limited by low stability. Despite that, alloying with relatively stable filled d-shell elements such as Au might help solve the stability issue, all while retaining good plasmonic properties. Moreover, s-block elements contribute only filled s-states and unfilled s- and p-states relevant to hot carrier generation, with a density of unfilled states increasing with energy above the Fermi level (Figure S2). Although DOS changes due to alloying are more complex than a simple linear combination of the constituent elements DOS, still this points to the possibility of increasing the proportion of higher energy unfilled states available for accepting an excited electron upon alloying.

While, as shown above, combining two filled d-shell elements results in a high energy tail of hot electrons, the tail is generally not skewed towards lower or higher energies. However, increasing the proportion of higher energy acceptor states points to the possibility of generating a higher proportion of more useful hot electrons, those at the very end of the high energy range (near the $E_{ph}$ above Fermi level).

To test this idea, as the model system we chose the $Na_8Au_4$ alloy of a defined crystal structure that has been experimentally observed.[39] These two elements do not form a simple FCC-like structure upon alloying so taking an approach of creating such structures, as was done in other, subsequently mentioned cases, would not be justified. Moreover, experimental observation of such a structure provides further validity to the choice of a model system.

As can be observed in Figure S2, the DOS of the alloy does show a skewed distribution of empty states. Furthermore, the d-bands of the alloy are still placed well below the Fermi level, and even moved further away from it. As a consequence, the hot carrier distribution is also skewed, resulting in an increasing probability of creating higher energy hot electrons for a relatively large range of photon energies, up to the appearance of low energy hot electron population coming from d-band contribution (Figure 5 - top panel). For illustration, the hot electron distribution of $Na_8Au_4$ alloy is compared to the more even hot carrier distribution of $Ag_5Au_3$ alloy at 2.8 eV photon energy on Figure 5 (bottom).

The results point to yet another way of hot carrier modulation by alloying. Once again, the hot carrier distribution can be rationalized by considering the electronic structure of the materials. They also suggest that by utilizing the knowledge of the changes that might occur using certain classes of elements as "ingredients", the resulting hot carrier distribution may be rationally engineered.

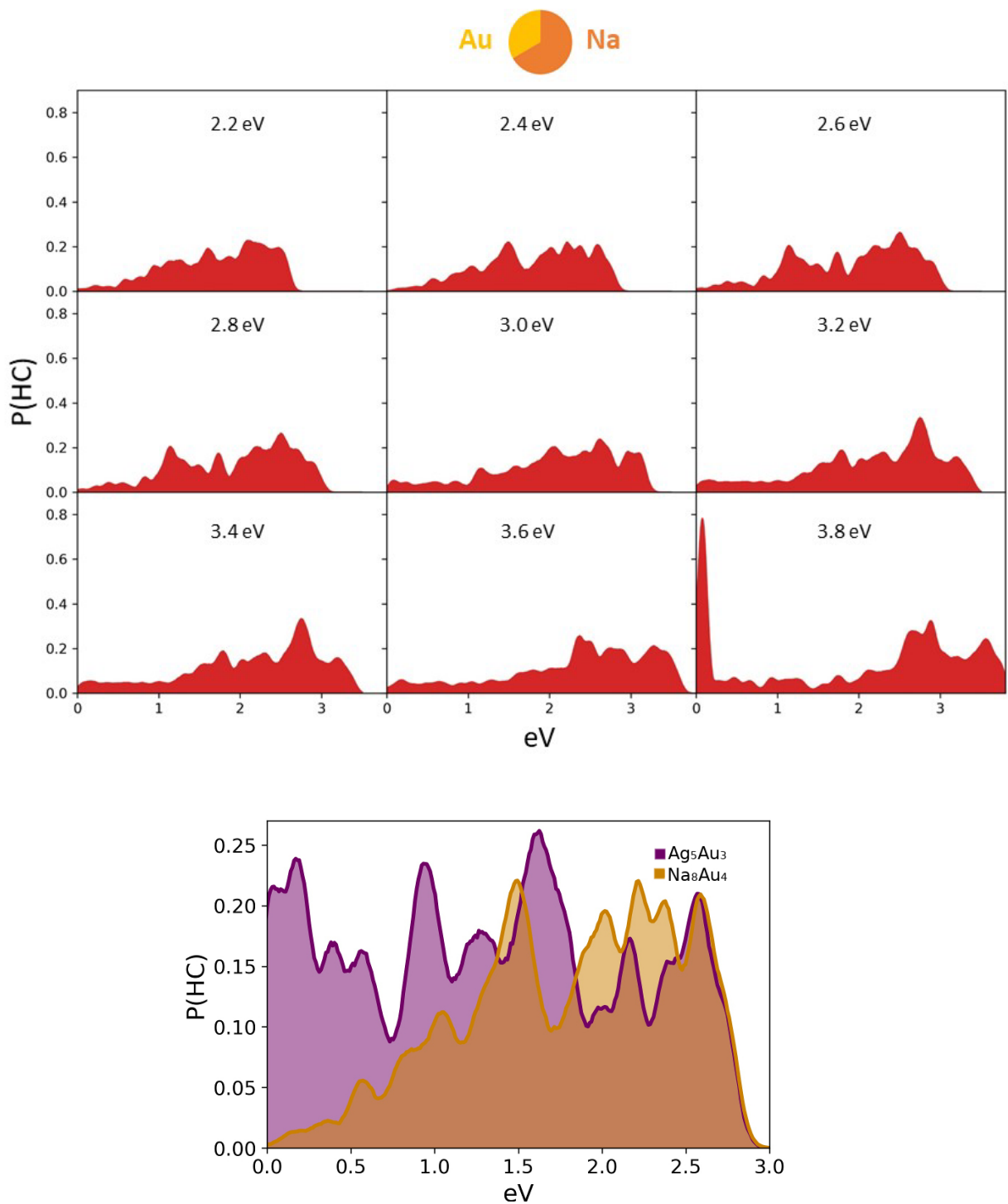

Figure 5: Hot electron energy distributions calculated for excitation of $Na_8Au_4$ alloy system with light ranging from 2.2 eV to 3.8 eV with a 0.2 eV increments (top). Comparison of hot electron distributions obtained by 2.8 eV light for $Na_8Au_4$ and $Ag_5Au_3$ systems (bottom).

# Influence of combining closed d-shell and p-block plasmonic metals - intermediate hot carrier generation with asymmetry tuning

Since Al is the most commonly used p-shell element in plasmonics, it is useful to study its hot carrier generation properties in alloys. Alloying Al with a filled d-shell element such as Cu leads to an FCC-like structure only in a very small composition range.[40] In this case we use a defined crystal structure available from experimental data.[41] It is visible from the Figure 6 that the resulting hot carrier distribution is more abundant than in case of mixing closed d-shell elements but less abundant than in case of open and closed d-shell elements mixing. The more abundant hot carrier generation can not, in this case, be ascribed to substantial density of high-lying d-states, but probably to higher relative density of sp-states. Furthermore, it is visible that the d-bands of Cu become lower upon alloying with Al (Figure S3). The alloy thus inherits the potential for asymmetry of hot hole and hot electron distribution that is characteristic for closed d-shell elements, with the possibility of tuning said asymmetry by the addition of Al.

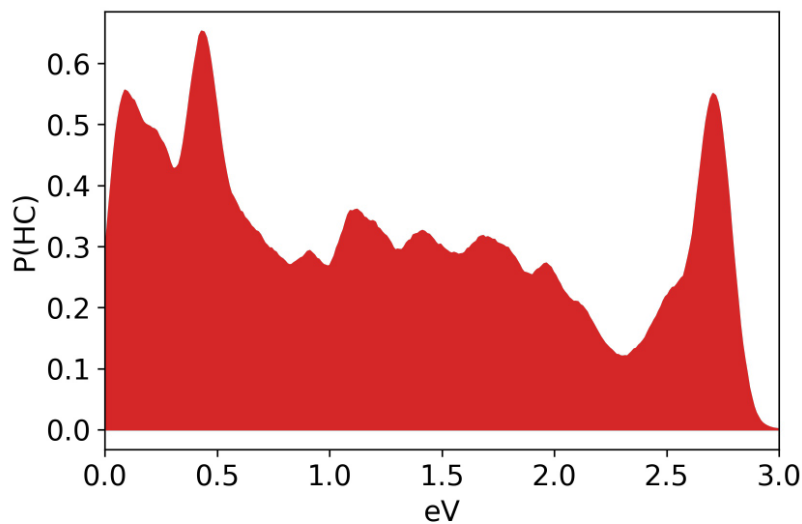

Figure 6: Hot electron energy distribution for $Al_2Cu_6$ alloy upon excitation with 2.8 eV light.

# Influence of combining open d-shell and p-block plasmonic metals - intermediate hot carrier generation with asymmetry tuning

Finally, the mixing of open d-shell and p-block elements is considered by investigating the $Al_3Pd_2$ system[42] (Figure 7). The hot carrier generation probability in $Al_3Pd_2$ alloy reaches similar (intermediate) values as in $Al_2Cu_6$, but skews towards higher energies resembling $Na_8Au_4$ in shape. This points to the tunability of both shape and magnitude of hot carrier energy distribution by pairing the right elements and perhaps classes of elements (based on their electronic structure). Furthermore, similarly to the case of Au alloyng with Al, the d-bands of Pd also lie much lower in the alloy with Al (more than 3 eV below the Fermi level as seen on Figure S4) than in pure Pd where they cross Fermi level. Thus alloying Al with unfilled d-shell elements provides exceptional tunability of the hot hole and hot electron distribution asymmetry.

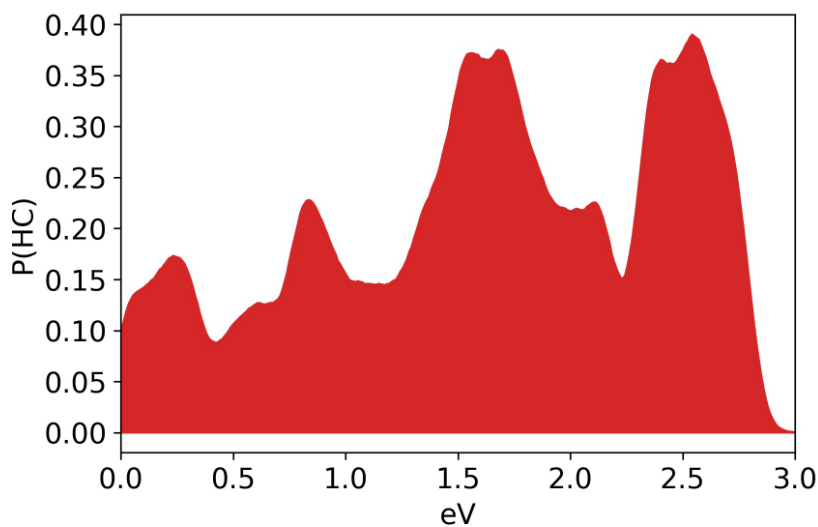

Figure 7: Hot electron energy distribution for $Al_3Pd_2$ alloy upon excitation with 2.8 eV light.

# Conclusions

In this paper we used a DFT-based method to study the effects of alloying on initial hot carrier energy distribution. The metals were divided into classes based on their electronic structure: closed d-shell elements (Au, Ag, Cu), open d-shell elements (Pd), p-shell elements (Al) and s-shell elements (Na).

We showed how alloying two closed d-shell elements can be used to avoid or introduce the abundant production of high energy hot holes, and how to tune it. Additionally, we observed that alloying unlocks new interband transitions which allow hot carrier production for all photon energies, and also lead to production of highly energetic hot electrons. Moreover, these interband transitions can be excited even with IR light of much lower energy than the nominal d-sp interband transition onset. We show that these effects are not a property of constituent metals in their pure form and are thus an emergent property of alloying. Since a method that neglects conservation of momentum would not be able to identify these emergent properties, we highlight the importance of taking conservation of momentum into account when studying the effects of alloying on hot carrier production.

Further, we showed that alloying a closed d-shell element with an open d-shell element modulates hot carrier generation by allowing an abundant production of HC with energies continuously ranging from low up to the photon energy. However, while it can be useful, this also means that a very efficient plasmon decay pathway is introduced, which can lead to a trade-off in hot carrier generation by reducing the field enhancement.

We also studied alloying s-block elements, which have recently been highlighted as prospective alternative plasmonic metals, with standard plasmonic closed d-shell metals. As a result of alloying, there is a rising density of states with increasing energy above the Fermi level. This combination can lead to a skewed hot carrier distribution, favoring the production of hot electrons at high energies. A larger proportion of high energy hot electrons can be useful both for purposes that require: i) only very high energy electrons such as injection over Schottky barrier or into an energetic unfilled molecular orbital, and ii) only require lower energy hot electrons, as they scatter and multiply into many lower energy hot electrons.

Finally, we studied alloying of p-shell elements, with open and closed d-shell elements. In both cases alloying results in moderate hot carrier generation probability, and point to tunability of the asymmetry of hot carrier distribution by matching such elements in an alloy.

As a step toward deliberate design of alloys that provide hot carrier distribution tailored for desired purposes, we analyzed the changes in the band structure and density of states that arise due to alloying, and connected them to the corresponding features in hot carrier distribution. This knowledge not only provides understanding of the observed changes, but also provides tools to match elements in an alloy knowing how it will tune the hot carrier distribution, thus enabling a conceptual framework for hot carrier engineering.

# Acknowledgement

The authors thank the financial support of the Croatian Science Foundation through the grant number IP-2019-04-5424.

# Supporting Information Available

Complementary data showing the effects of configuration on hot carrier distribution as well as density of states for different alloys are available in the supplementary information.

# References

(1) Tang, H.; Huang, C.-J. C. Z.; Bright, J.; Meng, G.; Liu, R.-S.; Wu, N. Plasmonic hot electrons for sensing, photodetection, and solar energy applications: A perspective. *The Journal of Chemical Physics* **2020**, *152*, 220901.

(2) Zhu, Y.; Xu, H.; Yu, P.; Wang, Z. Engineering plasmonic hot carrier dynamics toward efficient photodetection. *Applied Physics Reviews* **2021**, *8*, 021305.

(3) Zheng, B. Y.; Zhao, H.; Manjavacas, A.; McClain, M.; Nordlander, P.; Halas, N. J. Distinguishing between plasmon-induced and photoexcited carriers in a device geometry. *Nature Communications* **2015**, *6*, 7797.

(4) Su, Z.-C.; Chang, C.-H.; Jhou, J.-C.; Lin, H.-T.; Lin, C.-F. Ultra-thin Ag/Si heterojunction hot-carrier photovoltaic conversion Schottky devices for harvesting solar energy at wavelength above 1.1 µm. *Scientific Reports* **2023**, *13*, 5388.

(5) Ho, K. H. W.; Shang, A.; Shi, F.; Lo, T. W.; Yeung, P. H.; Yu, Y. S.; Zhang, X.; yin Wong, K.; Lei, D. Y. Plasmonic Au/TiO2-Dumbbell-On-Film Nanocavities for High-Efficiency Hot-Carrier Generation and Extraction. *Advanced Functional Materials* **2018**, *28*, 1800383.

(6) Zhang, D.; Choy, W. C. H.; Xie, F.; Sha, W. E. I.; Li, X.; Ding, B.; Zhang, K.; Huang, F.; Cao, Y. Plasmonic Electrically Functionalized TiO2 for High-Performance Organic Solar Cells. *Advanced Functional Materials* **2013**, *23*, 4255–4261.

(7) Liu, D.; Yang, D.; Gao, Y.; Ma, J.; Long, D. R.; Wang, D. C.; Xiong, P. Y. Flexible Near-Infrared Photovoltaic Devices Based on Plasmonic Hot-Electron Injection into Silicon Nanowire Arrays. *Angewandte Chemie International Edition* **2016**, *55*, 4577–4581.

(8) Carretero-Palacios, S.; Jiménez-Solano, A.; Míguez, H. Plasmonic Nanoparticles as Light-Harvesting Enhancers in Perovskite Solar Cells: A User's Guide. *ACS Energy Letters* **2016**, *1*, 323–331.

(9) Sarina, S.; Zhu, H.-Y.; Xiao, Q.; Jaatinen, E.; Jia, J.; Huang, Y.; Zheng, Z.; Wu, H. Viable Photocatalysts under Solar-Spectrum Irradiation: Nonplasmonic Metal Nanoparticles. *Angewandte Chemie* **2014**, *126*, 2979–2984.

(10) Halas, N. J. Introductory lecture: Hot-electron science and microscopic processes in plasmonics and catalysis. *Faraday Discussions* **2019**, *214*, 13–33.

(11) Lorber, K.; Zavašnik, J.; Sancho-Parramon, J.; Bubaš, M.; Mazaj, M.; Djinović, P. On the mechanism of visible-light accelerated methane dry reforming reaction over Ni/CeO$_{2-x}$ catalysts. *Applied Catalysis B: Environmental* **2022**, *301*, 120745.

(12) Monyoncho, E. A.; Dasog, M. Photocatalytic Plasmon-Enhanced Nitrogen Reduction to Ammonia. *Advanced Energy and Sustainability Research* **2021**, *2*, 2000055.

(13) Forno, S. D.; Ranno, L.; Lischner, J. Material, Size, and Environment Dependence of Plasmon-Induced Hot Carriers in Metallic Nanoparticles. *Journal of Physical Chemistry C* **2018**, *122*, 8517–8527.

(14) Wang, M.; Ye, M.; Iocozzia, J.; Lin, C.; Lin, Z. Plasmon-Mediated Solar Energy Conversion via Photocatalysis in Noble Metal/Semiconductor Composites. *Advanced Science* **2016**, *3*, 1600024.

(15) Abouelela, M. M.; Kawamura, G.; Matsuda, A. A review on plasmonic nanoparticle-semiconductor photocatalysts for water splitting. *Journal of Cleaner Production* **2021**, *294*, 126200.

(16) Reddy, H.; Wang, K.; Kudyshev, Z.; Zhu, L.; Yan, S.; Vezzoli, A.; Higgins, S. J.; Gavini, V.; Boltasseva, A.; Reddy, P. et al. Determining plasmonic hot-carrier energy distributions via single-molecule transport measurements. *Science* **2020**, *369*, 423–426.

(17) Rossi, T. P.; Erhart, P.; Kuisma, M. Hot-Carrier Generation in Plasmonic Nanoparticles: The Importance of Atomic Structure. *ACS Nano* **2020**, *14*, 9963–9971.

(18) Brown, A. M.; Sundararaman, R.; Narang, P.; Goddard, W. A.; Atwater, H. A. Nonradiative Plasmon Decay and Hot Carrier Dynamics: Effects of Phonons, Surfaces, and Geometry. *ACS Nano* **2016**, *10*, 957–966.

(19) Douglas-Gallardo, O. A.; Berdakin, M.; Frauenheim, T.; Sánchez, C. G. Plasmon-induced hot-carrier generation differences in gold and silver nanoclusters. *Nanoscale* **2019**, *11*, 8604–8615.

(20) Jin, H.; Kahk, J. M.; Papaconstantopoulos, D. A.; Ferreira, A.; ; Lischner, J. Plasmon-Induced Hot Carriers from Interband and Intraband Transitions in Large Noble Metal Nanoparticles. *PRX Energy* **2022**, *1*, 013006.

(21) Bubaš, M.; Sancho-Parramon, J. DFT-Based Approach Enables Deliberate Tuning of Alloy Nanostructure Plasmonic Properties. *Jorunal of Physical Chemistry C* **2021**, *125*, 24032–24042.

(22) Mortensen, J. J.; Hansen, L. B.; Jacobsen, K. W. Real-space grid implementation of the projector augmented wave method. *Physical Review B* **2005**, *71*, 035109.

(23) Enkovaara, J.; Rostgaard, C.; Mortensen, J. J.; Chen, J.; Dułak, M.; Ferrighi, L.; Gavnholt, J.; Glinsvad, C.; Haikola, V.; Hansen, H. A. et al. Electronic structure calculations with GPAW: a real-space implementation of the projector augmented-wave method. *Journal of Physics: Condensed Matter* **2010**, *22*, 253202.

(24) Larsen, A. H.; Mortensen, J. J.; Blomqvist, J.; Castelli, I. E.; Christensen, R.; Dułak, M.; Friis, J.; Groves, M. N.; Hammer, B.; Hargus, C. et al. The atomic simulation environment—a Python library for working with atoms. *Journal of Physics: Condensed Matter* **2017**, *29*, 273002.

(25) Perdew, J. P.; Ruzsinszky, A.; Csonka, G. I.; Vydrov, O. A.; Scuseria, G. E.; Constantin, L. A.; Zhou, X.; Burke, K. Restoring the Density-Gradient Expansion for Exchange in Solids and Surfaces. *Physical Review Letters* **1976**, *100*, 136406.

(26) White, T. P.; Catchpole, K. R. Plasmon-enhanced internal photoemission for photovoltaics: Theoretical efficiency limits. *Applied Physics Letters* **2012**, *101*, 073905.

(27) Gong, T.; Munday, J. N. Materials for hot carrier plasmonics. *Optical Materials Express* **2015**, *5*, 2501–2512.

(28) Krayer, L. J.; Palm, K. J.; Gong, C.; Torres, A.; Villegas, C. E.; Rocha, A. R.; Leite, M. S.; Munday, J. N. Enhanced near-infrared photoresponse from nanoscale Ag-Au alloyed films. *ACS Photonics* **2020**, *7*, 1689–1698.

(29) Smith, N. V. Photoelectron energy spectra and the band structures of the noble metals. *Physical Review B* **1971**, *3*, 1862.

(30) Sundararaman, R.; Narang, P.; Jermyn, A. S.; III, W. A. G.; Atwater, H. A. Theoretical predictions for hot-carrier generation from surface plasmon decay. *Nature Communications* **2014**, *5*, 5788.

(31) Alex Zunger, L. G. F., S.-H. Wei; Bernard, J. E. Special quasirandom structures. *Physical Review Letters* **1990**, *65*, 353.

(32) Gong, C.; Kaplan, A.; Benson, Z. A.; Baker, D. R.; McClure, J. P.; Rocha, A. R.; Leite, M. S. Band Structure Engineering by Alloying for Photonics. *Advanced Optical Materials* **2018**, *6*, 1802185.

(33) Liu, J. G.; Zhang, H.; Link, S.; Nordlander, P. Relaxation of Plasmon-Induced Hot Carriers. *ACS Photonics* **2018**, *5*, 2584–2595.

(34) Stofela, S. K.; Kizilkaya, O.; Diroll, B. T.; Leite, T. R.; Taheri, M. M.; Willis, D. E.; Baxter, J. B.; Shelton, W. A.; Sprunger, P. T.; McPeak, K. M. A Noble-Transition Alloy Excels at Hot-Carrier Generation in the Near Infrared. *Advanced Materials* **2020**, *32*, 1906478.

(35) Rahm, J. M.; Tiburski, C.; Rossi, T. P.; Nugroho, F. A. A.; Nilsson, S.; Langhammer, C.; Erhart, P. A Library of Late Transition Metal Alloy Dielectric Functions for Nanophotonic Applications. *Advanced Functional Materials* **2020**, *30*, 2002122.

(36) Sánchez, C. G.; Berdakin, M. Plasmon-Induced Hot Carriers: An Atomistic Perspective of the First Tens of Femtoseconds. *Journal of Physical Chemistry C* **2022**, *126*, 10015–10023.

(37) Hopper, E. R.; Boukouvala, C.; Asselin, J.; Biggins, J. S.; Ringe, E. Opportunities and Challenges for Alternative Nanoplasmonic Metals: Magnesium and Beyond. *Journal of Physical Chemistry C* **2022**, *126*, 10630–10643.

(38) Kuisma, M.; Rousseaux, B.; Czajkowski, K. M.; Rossi, T. P.; Shegai, T.; Erhart, P.; Antosiewicz, T. J. Ultrastrong Coupling of a Single Molecule to a Plasmonic Nanocavity: A First-Principles Study. *ACS Photonics* **2022**, *9*, 1065–1077.

(39) Haucke, W. Ueber gold-natrium-legierungen. *Naturwissenschaften* **1937**, *25*, 61–61.

(40) Shahcheraghi, N.; Keast, V.; Gentle, A.; Arnold, M.; Cortie, M. Anomalously strong plasmon resonances in aluminium bronze by modification of the electronic density-of-states. *Journal of Physics: Condensed Matter* **2016**, *28*, 405501.

(41) Kurdjumov, G.; Mireckij, V.; Stelleckaja, T. Transformations in eutectoid alloys of cu - al. v. structure of the martensitic phase gamma' and the mechanism of the beta1 - gamma' transformation. *Zhurnal Tekhnicheskoi Fiziki* **1938**, *8*, 1959–1972.

(42) M.Ellner; Kattner, U.; Predel, B. Konstitutinelle und strukturelle untersuchungen im system pd-al. *Journal of the Less-Common Metals* **1982**, *87*, 117–133.